\documentclass[twocolumn,preprintnumbers,amssymb,aps,pra]{revtex4}
\usepackage{graphicx}
\usepackage{dcolumn}
\usepackage{bm}
\usepackage{latexsym,epsfig}

\usepackage{array}
\usepackage{color}
\usepackage{ulem}
\usepackage{amstext}

\begin{document}

\title{Roles of electron correlation effects for accurate determination of $g_j$ factors of low-lying states of $^{113}$Cd$^+$ and their applications to atomic clock}
\vspace*{0.5cm}

\author{J. Z. Han$^{1,5}$, Y. M. Yu$^2$\footnote{Email: ymyu@aphy.iphy.ac.cn}, B. K. Sahoo$^3$\footnote{Email: bijaya@prl.res.in},
J. W. Zhang$^{4,5}$\footnote{Email: zhangjw@tsinghua.edu.cn} and L. J. Wang$^{1,4,5}$}

\affiliation{$^1$Department of Physics, Tsinghua University, Beijing 100084, China}
\affiliation{$^2$Beijing National Laboratory for Condensed Matter Physics, Institute of Physics, Chinese Academy of Sciences, Beijing 100190, China}
\affiliation{$^3$Atomic, Molecular and Optical Physics Division, Physical Research Laboratory, Navrangpura, Ahmedabad-380009, India}
\affiliation{$^4$Department of Precision Instruments, Tsinghua University, Beijing 100084, China}
\affiliation{$^5$State Key Laboratory of Precision Measurement Technology and Instruments, Tsinghua University, Beijing 100084, China}

\date{Received date; Accepted date}

\vskip1.0cm

\begin{abstract}
We investigate roles of electron correlation effects in the determination of $g_j$ factors of the $ns~^2S_{1/2}$ ($n$=5,6,7),
$np~^2P_{1/2,3/2}$ ($n$=5,6), $5d~^2D_{3/2,5/2}$, and $4f~^2F_{5/2,7/2}$ states of the singly ionized cadmium (Cd$^+$) ion. Single and
double excited configurations along with important valence triple excited configurations through relativistic coupled-cluster (RCC)
theory are taken into account for incorporating electron correlation effects in our calculations. We find significant contributions
from the triples to the lower $S$ and $P$ states for attaining high accuracy results. The contributions of Breit interaction and lower-order
quantum electrodynamics effects, such as vacuum polarization and self-energy corrections, are also estimated using
the RCC theory and are quoted explicitly. In addition, we present energies of the aforementioned states from our calculations and compare
them with the  experimental results to validate $g_j$ values. Using the $g_j$ factor of the ground state, systematical shift due to the
Zeeman effect in the microwave clock frequency of the $|5s~^2S_{1/2}, F=0,m_F=0 \rangle \leftrightarrow |5s~^2S_{1/2}, F=1,m_F=0 \rangle$
transition in $^{113}$Cd$^+$ ion has been estimated.
\end{abstract}

\maketitle

\section{Introduction}

Urge of high precision atomic clocks for both scientific and commercial applications are well acknowledged in many areas of physics. Today
optical atomic clocks offer the most precise frequencies to recalibrate unit of time (i.e. second). Howbeit, use of microwave atomic clocks
are extremely robust in numerous scientific and industrial fields including satellite navigation systems, network synchronization, timekeeping
applications and defense systems etc. \cite{Weyers-MT-2018} owing to their several order lower frequencies than the optical clock frequencies.
Mostly microwave clocks are based on the neutral atoms, but making these clocks using singly charged ions have many advantages. They can be
more compact in size and consume low power, that are the imperative criteria for making a portable atomic clock. Compared with the other
developed trapped ion microwave clocks, e.g. $^{199}$Hg$^+$ \cite{Phoonthong-APB-2014,Burt-trans-2016}, microwave clock using $^{113}$Cd$^{+}$ has a unique feature as its cooling and pumping lines have the frequency difference of only 800 MHz. Therefore, its cooling, pumping and detecting processes can be carried out by the same laser. This is advantageous for making miniaturized atomic clock for aerospace applications \cite{Zhangjw-PRA-2012}. Since 2012, we have progressed a lot in making high-performance miniaturized cadmium ion atomic clocks
\cite{Zhangjw-PRA-2012, Wangsg-OE-2013,Zhangjw-APB-2014,Miaok-OL-2015}. We achieved the frequency uncertainty and stability for the ground
state hyperfine splitting to $6.6 \times 10^{-14}$ and $6.1 \times 10^{-13}/\sqrt{\tau}$, where $\tau$ is the average time, respectively,
in 2015 \cite{Miaok-OL-2015}. We have accomplished sympathetic cooling of Cd$^+$ ions in the mean time, which lowers the second-order
Doppler shift and dead-time drastically \cite{Zuoyn-arxiv-2019}. In order to improve uncertainty of the clock frequency to below $10^{-16}$
or making comparable with the currently available $^{133}$Cs fountain clocks, one of the important tasks for us is to estimate the
second-order Zeeman shift accurately.

In the microwave atomic clocks a small external magnetic field is applied to break the degeneracy of the ground state hyperfine
levels. The fluctuation arising from the background magnetic field can be suppressed down to $10^{-9}$ Tesla or even lower by using
multi-layer magnetic shields. When the stability of the background magnetic field is well controlled, it becomes important to calibrate the applied
magnetic field very strictly. This can affect precise estimate of the Zeeman shift, hence accurate determination of the clock transition.
For this purpose precise knowledge of the $g_j$ factor of the ground state of Cd$^+$ ion is strongly desired. There has also been immense
interest to understand roles of various physical effects for the accurate determination of the $g_j$ factors of atomic states. Most of the
$g_j$ factor studies are concentrated on the highly charged ions (HCIs) with few electrons in which relativistic effects play the crucial
roles for their accurate determination. It demands for rigorous treatments of quantum electrodynamics (QED) to higher-orders and nuclear
recoil (NR) effects. For example, agreement between the theoretical and experimental values of the $g_j$ factors in the H-like C$^{5+}$, O$^{7+}$,
and Si$^{13+}$ ions \cite{Verdu-PRL-2004,Sturm-PRA-2013,Sturm-nature-2014,Kohler-JPB-2015}, and in the Li-like Si$^{11+}$ and Ca$^{17+}$
\cite{Wagner-PRL-2013,Volotka-PRL-2014,Kohler-ncom-2016} HCIs at the 8$^{th}$ or even at the lower decimal places serve as the most stringent test
of the bound-state QED theory. Such studies yield unprecedented values of the ratio between the mass of an electron and the mass of a proton,
and the fine structure constant \cite{Sturm-atoms-2017}. Contrasting to the great success in achieving high precision values of the $g_j$
factors in few-body systems, accurate determination of these factors in many-electron systems is challenging owing to strong electron
correlation effects associated with this property. In the neutral atoms or singly charged ions, these interactions contribute predominantly
to the $g_j$ factors over the QED interactions \cite{Veseth-PRA-1980,Veseth-JPB-1983,Dzuba-PS-1985,Gossel-PRA-2013}. In this context, the
relativistic coupled-cluster (RCC) theory, which is currently known as one of the leading quantum many-body methods and has been referred
to as the gold standard for treating electron correlations, is apt to determine atomic properties including $g_j$ factors accurately. This
theory was employed to study the $g_j$ factors of the ground states of Li, Be$^+$, and Ba$^+$ by Lindroth and Ynnerman
\cite{Lindroth-PRA-1993}, but they had determined only the corrections to the $g_j$ factors ($\Delta g_j$) due to electron correlation
effects with respect to the bare Dirac values. Recently, the roles of electron correlation effects to the net $g_j$ factors of ground and
few excited states of Ca$^+$ were demonstrated by employing RCC theory \cite{SahooKumar-PRA-2017}.

In this work, we have applied the RCC theory to calculate the $g_j$ factors of the ground state $5s~^2S_{1/2}$ and some of the low-lying
$ns~^2S_{1/2}$ ($n$=6,7), $np~^2P_{1/2,3/2}$ ($n$=5,6), $5d~^2D_{3/2,5/2}$, and $4f~^2F_{5/2,7/2}$ excited states of Cd$^+$. We have also
determined electron affinities (EAs) of the valence electrons of the above states with the $[4d^{10}]$ closed-shell configuration by considering
singles and doubles excitations approximation in the RCC theory (RCCSD method) and compared them with the available experimental results to
validate our calculations. We have incorporated contributions from the important valence triples excitations in a perturbative approach in
the RCCSD method (RCCSDpT method) only in the $g_j$ factor evaluation expression as described in \cite{SahooKumar-PRA-2017}. Further,
systematic shift due to the Zeeman effect in the microwave clock frequency of $^{113}$Cd$^+$ ion has been estimated by using the calculated
$g_j$ factor of its ground state.

\section{Theory}

In the presence of an external homogeneous magnetic field ${\vec {\bf B}}$, the interaction Hamiltonian of electrons in an atomic system
is given by \cite{Sakurai1967}
\begin{eqnarray}
 H_{mag} &=& ec \sum_i \mbox{\boldmath${\vec \alpha}$}_i \cdot {\vec \textbf{A}}_i \nonumber \\
         &=& - \frac{ec}{2} \sum_i \mbox{\boldmath${\vec \alpha}$}_i \cdot ({\vec \textbf{r}}_i \times {\vec \textbf{B}} ),
\end{eqnarray}
where $e$ is the electric charge of the electron, $c$ is the speed of light, $\mbox{\boldmath${\vec \alpha}$}$ is the Dirac operator, and
${\vec \textbf{A}}$ is the vector field seen by the electron located at ${\vec {\bf r}}$ due to the applied magnetic field. We can rewrite the above expression as
\begin{eqnarray}
 H_{mag} &=& -\frac{ec}{2} \sum_i ( \mbox{\boldmath${\vec \alpha}$}_i \times {\vec \textbf{r}}_i) \cdot {\vec \textbf{B}}  \nonumber \\
 &=& i \frac{ec}{\sqrt{2}} \sum_i r_i \left \{ \mbox{\boldmath${\vec \alpha}$}_i \otimes {\vec \textbf{C}}^{(1)} \right \}^{(1)} \cdot {\vec {\textbf B}} \nonumber \\
 &=& {\bf \vec {\cal M}} \cdot {\vec {\textbf B}}
 \end{eqnarray}
where ${\vec \textbf{C}}^{(1)}$ is the Racah coefficient of rank one and magnetic moment operator given by
\begin{eqnarray}
{\vec {\cal M}} = \sum_{i, q=-1,0,1} \mbox{\boldmath${\vec \mu}$}_q^{(1)} (r_i)=
\sum_i r_i \left \{ \mbox{\boldmath${\vec \alpha}$}_i \otimes {\vec \textbf{C}}^{(1)} \right \}^{(1)}.
\end{eqnarray}
Using this operator, the Dirac contribution to the Lande $g_j$ factor of a bound-state electron in an atomic system is given by
\begin{equation}
g_j^D = -\frac{1}{\mu_B} \frac{{\bf \vec {\cal M}}}{ \vec \textbf{J}}
\end{equation}
with total angular momentum of the state $\textbf{J}$ and the Bohr magneton $\mu_B = e \hbar / 2m_e$ for mass of the electron $m_e$ and
Planck constant $\hbar$. Using the reduced matrix element, it can be expressed as
\begin{equation}
g_j^D = - \frac{1}{2 \mu_B } \frac{\langle J|| {\vec {\cal M}} ||J \rangle}{ \sqrt{J(J+1)(2J+1)}}.
 \label{eqn5}
\end{equation}
For the calculation of this factor, the single particle reduced matrix element of $\mbox{\boldmath${\mu}$}_q^{(1)}$ is given by
\begin{eqnarray}
\langle \kappa_{f}||\mbox{\boldmath${ \mu}$}^{(1)}||\kappa_{i}\rangle &=&-(\kappa_{f}+\kappa_{i})\langle -\kappa_{f}||\textbf{C}^{(1)}||\kappa_{i}\rangle \nonumber \\
&&\times\int^{\infty}_{0}dr \ r \ \left (P_{f}Q_{i}+Q_{f}P_{i} \right) ,
\label{eqn6}
\end{eqnarray}
where $P(r)$ and $Q(r)$ are the large and small components of the radial parts of the single particle Dirac orbitals, respectively,
and $\kappa$ is the relativistic angular momentum quantum number. The reduced matrix element of the Racah ${\vec \textbf{C}}^{(1)}$ operator is calculated as
\begin{eqnarray}
\langle \kappa_f\, ||\, \textbf{C}^{(k)}\,||\, \kappa_i \rangle &=& (-1)^{j_f+1/2} \sqrt{(2j_f+1)(2j_i+1)} \ \ \ \ \ \ \ \ \nonumber \\
                  &&  \left ( \begin{matrix} {
                          j_f & k & j_i \cr
                          1/2 & 0 & -1/2 \cr }
                         \end{matrix}
                            \right ) \Pi(l_{\kappa_f},k,l_{\kappa_i}), \ \ \ \ \
\end{eqnarray}
with
\begin{eqnarray}
\Pi(l_{\kappa_f},k,l_{\kappa_i}) &=&
\left\{\begin{array}{ll}
\displaystyle
1 & \mbox{for } l_{\kappa_f}+k+l_{\kappa_i}= \mbox{even}
\\ [2ex]
\displaystyle
0 & \mbox{otherwise,}
\end{array}\right.
\label{eqn12}
\end{eqnarray}
for the corresponding orbital momentum $l_{\kappa}$ of the orbital with relativistic quantum number $\kappa$.

\begin{table*}[t]
\caption{Electron affinities (EAs) of the $5-7s~^2S_{1/2}$, $5-6p~^2P_{1/2,3/2}$, $5d~^2D_{3/2,5/2}$, and $4f~^2F_{5/2,7/2}$ states
(in cm$^{-1}$) of $^{113}$Cd$^{+}$ from DHF, second-order relativistic many-body perturbation theory (RMBPT(2)) and RCCSD methods are given. Corrections
from the Breit ($\Delta$Breit) and low-order QED ($\Delta$QED) interactions are quoted separately. The final results are obtained by the
RCCSD values adding $\Delta$Breit and  $\Delta$QED are compared with the experimental values listed in the NIST database \cite{NIST}.
Uncertainties to our final results are estimated using triple excitations in the perturbative approach. Differences between our results
from the NIST data are mentioned as $\Delta$ in percentage.}
{\setlength{\tabcolsep}{9pt}
\begin{tabular}{lcccccccc}\hline\hline \\
State	 &	DHF	 &	RMBPT(2)	 &	RCCSD 	 &	$\Delta$Breit	 &$\Delta$QED&	Final	 &	 NIST 	 &	$\Delta $	\\\hline	\\										
$5s~^2S_{1/2}$ 	 &	-124567.93 	 &	-138618.14 	 &	-136012.37 	 &	84.10 	 &	60.14 	 &	-135868(587)	 &	-136374.34 	 &	0.4 	\\											
$5p~^2P_{1/2}$ 	 &	-84902.34 	 &	-93102.78 	 &	-91847.93 	 &	83.31 	 &	7.58 	 &	-91757(306)	 &	-92238.40 	 &	0.5 	\\											
$5p~^2P_{3/2}$ 	 &	-82870.41 	 &	-90363.14 	 &	-89347.42 	 &	50.24 	 &	-0.64 	 &	-89298(295)	 &	-89755.93 	 &	0.5 	\\											
$6s~^2S_{1/2}$ 	 &	-51061.62 	 &	-53896.08 	 &	-53315.36 	 &	18.02 	 &	12.42 	 &	-53285(147)	 &	-53383.93 	 &	0.2 	\\											
$5d~^2D_{3/2}$ 	 &	-45146.73 	 &	-46880.67 	 &	-46602.79 	 &	3.85 	 &	0.38 	 &	-46599(94)	 &	-46685.35 	 &	0.2 	\\											
$5d~^2D_{5/2}$ 	 &	-45009.55 	 &	-46667.87 	 &	-46445.57 	 &	-0.60 	 &	0.01 	 &	-46446(90)	 &	-46530.84 	 &	0.2 	\\											
$6p~^2P_{1/2}$ 	 &	-39865.05 	 &	-41935.34 	 &	-41536.83 	 &	22.69 	 &	1.98 	 &	-41512(78)	 &	-41664.22 	 &	0.4 	\\											
$6p~^2P_{3/2}$ 	 &	-39241.67 	 &	-41181.53 	 &	-40844.47 	 &	13.81 	 &	-0.25 	 &	-40831(75)	 &	-40990.98 	 &	0.4 	\\											
$7s~^2S_{1/2}$ 	 &	-28191.88 	 &	-29280.12 	 &	-29045.33 	 &	7.11 	 &	4.87 	 &	-29033(58)	 &	-29073.79 	 &	0.1 	\\											
$4f~^2F_{5/2}$ 	 &	-27539.36 	 &	-27884.18 	 &	-27880.85 	 &	0.13 	 &	0.11 	 &	-27881(22)	 &	-27955.19 	 &	0.3 	\\											
$4f~^2F_{7/2}$ 	 &	-27542.58 	 &	-27886.78 	 &	-27884.13 	 &	0.23 	 &	0.11 	 &	-27884(22)	 &	-27942.22 	 &	0.2 	\\ \hline\hline											
\end{tabular}}
\label{CdEA}
\end{table*}

For evaluating $g_j^D$ using Eq. (\ref{eqn5}), it is necessary to calculate wave functions of the states in an atomic system considering
relativistic effects. It is also known that the Dirac value of Lande $g$ factor of a free electron ($g_f^D$) has significant corrections
from the QED theory. The net value with the QED effects is approximately given by \cite{Czarnecki2005}
\begin{eqnarray}
 g_{f} & \simeq & g_f^D \times \left [ 1 + \frac{1}{2} \frac{\alpha_e}{\pi } - 0.328 \left ( \frac{\alpha_e}{\pi} \right )^2 + \cdots  \right ] \nonumber \\
   &\approx & 1.001160 \times g_f^D ,
 \end{eqnarray}
where $\alpha_e$ is the fine structure constant. To account for this correction along with $g_j^D$ (denoted by $\Delta g_j^Q$) for
the net result as $g_j=g_j^D + \Delta g_j^Q$, we consider the additional interaction Hamiltonian with the magnetic field
as \cite{Akhiezer1965}
\begin{eqnarray}
 \Delta H_{mag}  & \approx & 0.001160 \ \mu_B \beta \ \mbox{\boldmath${ \vec \Sigma}$} \cdot {\vec \textbf{B}} \nonumber \\
  &=& 0.001160 \ \Delta {\bf \vec {\cal M}} \cdot {\vec {\textbf B}} ,
\end{eqnarray}
where $\beta$ is the Dirac matrix, $\mbox{\boldmath${ \vec \Sigma}$}$ is the four-component spinor and ${\Delta \vec {\cal M}} = \sum_{i, q=-1,0,1} \Delta
\mbox{\boldmath${\vec \mu}$}_q^{(1)} (r_i)=\sum_i \beta_i \mbox{\boldmath${ \vec \Sigma}$}_i$. Using this Hamiltonian, we determine
$\Delta g_j^Q$ as \cite{Cheng1985}
\begin{equation}
\Delta g_j^Q =0.001160 \ \frac{\langle J|| \Delta {\bf \vec {\cal M}} ||J \rangle}{\sqrt{J(J+1)(2J+1)}},
\label{BQED}
\end{equation}
for which the single particle matrix element is given by
\begin{eqnarray}
\langle \kappa_{f}||\Delta \mbox{\boldmath${ \mu}$}^{(1)} || \kappa_{i}\rangle &=& (\kappa_f + \kappa_i -1)
  \langle - \kappa_{f}||\textbf{C}^{(1)}||\kappa_{i} \rangle \nonumber \\ && \times \int^{\infty}_{0}dr
(P_{f}P_{i}+Q_{f}Q_{i}).
\label{eqn8}
\end{eqnarray}

Contribution due to the NR effect to the bound state $g_j$ factors in Cd$^+$ can be estimated using the formula \cite{Shabaev-JPCRD-2015}
\begin{equation}
\Delta g_j^{\rm{NR}}=\frac{(\alpha Z)^2}{n^2}\frac{m}{M},
\end{equation}
where $m$ and $M$ are masses of an electron and atomic nucleus, $Z$ is the atomic number, and $n$ is the principal quantum number of the
interested states. This is found to be of the order of $\sim 10^{-7}$; which is neglected because such uncertainty is much below than the intended precision
that can be achieved in the present work.

\section{Method of calculations}\label{sec3}

We consider the Dirac-Coulomb (DC) Hamiltonian to calculate the wave functions of the atomic states, which is given in atomic unit (a.u.) by
\begin{equation}\label{eq:DHB}
H^{DC} =\sum_i \left [c\mbox{\boldmath$\alpha$}_i\cdot \textbf{p}_i+(\beta_i-1)c^2+V_n(r_i)\right]+\sum_{i,j>i}\frac{1}{r_{ij}},\\
\end{equation}
where $\textbf{p}_i$ is the momentum operator, $V_n(r)$ denotes the nuclear potential, and $\frac{1}{r_{ij}}$
represents for the Coulomb potential between the electrons located at the $i$ and $j$ positions. The Breit interaction contribution is
estimated by incorporating the interaction potential
\begin{equation}\label{eq:DHB}
V^B(r_{ij}) = - \frac{[\mbox{\boldmath$\alpha$}_i\cdot\mbox{\boldmath$\alpha$}_j+
(\mbox{\boldmath$\alpha$}_i\cdot\mathbf{\hat{r}_{ij}})(\mbox{\boldmath$\alpha$}_j\cdot\mathbf{\hat{r}_{ij}})]}{2r_{ij}} ,
\end{equation}
where $\mathbf{\hat{r}_{ij}}$ is the unit vector along $\mathbf{r_{ij}}$. Similarly, we also include effective potentials for vacuum
polarization (VP) and self-energy (SE) interactions as discussed in our previous work \cite{Yuym-PRA-2019} to account for QED interactions
in the determination of the atomic wave functions.

The investigated states of Cd$^{+}$ ion can be expressed in the RCC theory as \cite{Yuym2017,Licb2018}
\begin{equation}
 \vert \Psi_v \rangle = e^T \{ 1+ S_v \} \vert \Phi_v \rangle,
\end{equation}
where $|\Phi_v\rangle=a^\dag_v|\Phi_0\rangle$ is the reference state with valence orbital $v$ for the Dirac-Hartree-Fock (DHF) wave function
of the $[4s^24p^64d^{10}]$ closed-shell configuration. The RCC excitation operators $T$ and $S_v$ are responsible for exciting electrons from
the $|\Phi_0\rangle$ and $|\Phi_v\rangle$ reference states, respectively. The amplitudes of these RCC operators are evaluated by solving the
following equations
\begin{equation}
 \langle \Phi_0^* \vert \overline{H}_N  \vert \Phi_0 \rangle = 0
\label{eqt}
\end{equation}
and
\begin{equation}
 \langle \Phi_v^* \vert \big ( \overline{H}_N - \Delta E_v \big ) S_v \vert \Phi_v \rangle =  - \langle \Phi_v^* \vert \overline{H}_N \vert \Phi_v \rangle ,
\label{eqsv}
 \end{equation}
where $|\Phi_0^* \rangle$ and $|\Phi_v^* \rangle$ are the singly and doubly excited state configuration with respect to
$|\Phi_0 \rangle$ and $|\Phi_v \rangle$, respectively. The notation $\overline{H}_N$ is defined as $\overline{H}_N=(He^T)_l$ with
subscript $N$ means normal order form of the operator and $l$ means that all the terms are linked. The quantity $\Delta E_v$ corresponds to
EA of the state with the valence electron $v$. We evaluate $\Delta E_v$ by
\begin{equation}
 \Delta E_v=\langle\Phi_v |\overline{H}_N \left \{ 1+S_v \right \}| \Phi_v \rangle-\langle\Phi_0|\overline{H}_N|\Phi_0\rangle .
 \label{eqeng}
\end{equation}
In the RCCSD method, the singles and doubles excitations are denoted by
\begin{eqnarray}
T=T_1+T_2  \ \ \ \text{and} \ \ \ S_v=S_{1v}+S_{2v} .
\label{eq:tsv}
\end{eqnarray}

After obtaining amplitudes of the RCC operators, the expectation value of an operator $O$ is evaluated as
\begin{eqnarray}
\frac{\langle \Psi_v \vert O \vert \Psi_v \rangle}{ \langle \Psi_v \vert \Psi_v\rangle }
&=& \frac{\langle \Phi_v | \{1+ S_v^{\dagger}\} e^{T^{\dagger}} O e^T \{1+S_v \} | \Phi_v \rangle} {\langle \Phi_v | \{1+ S_v^{\dagger}\}
e^{T^{\dagger}} e^T \{1+S_v \} | \Phi_v \rangle } . \ \ \ \
\label{prpeq}
\end{eqnarray}
We adopt an iterative procedure to include contributions from the non-terminative terms from the above expressions. Here, $O$ stands for both
the ${\bf {\cal M}}$ and $\Delta {\bf {\cal M}}$ operators for the evaluations of the $g_j^D$ and $\Delta g_j^Q$ contributions, respectively.
In our previous work, we had observed that triples excitations play important roles in the determination of the $g_j$ factors
\cite{SahooKumar-PRA-2017}. Inclusion of these excitations require huge computational resources, which we are lacking at present. Therefore,
we take into account these contributions in the RCCSDpT method only by defining excitation operators in the perturbative approach as
\begin{equation}
 T_{3}^{\rm{pert}}  = \frac{1}{6} \sum_{abc,pqr} \frac{\big ( H_N T_2 \big )_{abc}^{pqr}}{\epsilon_a + \epsilon_b + \epsilon_c - \epsilon_p -\epsilon_q - \epsilon_r}
\label{t3eq}
 \end{equation}
and
\begin{equation}
 S_{3v}^{\rm{pert}} = \frac{1}{4} \sum_{ab,pqr} \frac{\big ( H_N T_2 + H_N S_{2v} \big )_{abv}^{pqr}}{\Delta E_v + \epsilon_a + \epsilon_b - \epsilon_p -\epsilon_q - \epsilon_r} ,
\label{s3eq}
 \end{equation}
where $\{a,b,c \}$ and $\{ p,q,r \}$ represent for the occupied and virtual orbitals, respectively, and $\epsilon$s are their single particle
orbital energies. Contributions from these operators to the $g_j$ factors are then given in terms of $T_2^{\dagger}OT_{3}^{pert}$,
$S_{2v}^{\dagger}OT_{3}^{pert}$, $S_{2v}^{\dagger}OS_{3v}^{pert}$, $T_2^{\dagger}OS_{3v}^{pert}$, $S_{1v}^{\dagger}T_2^{\dagger}OS_{3v}^{pert}$,
$T_{3}^{pert \dagger}OT_{3}^{pert}$, and $S_{3v}^{pert \dagger}OS_{3v}^{pert}$ along with their possible complex conjugate (c.c.) terms
as part of Eq. (\ref{prpeq}) in the RCCSD expression.

\section{Results and Discussion}

To verify accuracies of the wave functions of the atomic states, whose $g_j$ factors are investigated here, we have evaluated EAs and
compared them with their available experimental values. In Table \ref{CdEA}, we give EAs of 11 low-lying states of Cd$^+$ from the DHF,
second-order relativistic many-body perturbation theory (RMBPT(2)) and RCCSD methods. As can be seen, the DHF method gives lower values
while the RMBPT(2) method gives larger values compared to the results from the RCCSD method. The RCCSD results are in close agreement with
the experimental results. Corrections from the Breit and QED interactions are quoted explicitly from the RCCSD method, and those corrections
are found to be comparatively small. Uncertainties to our final results are given by estimating contributions from the valence triple
excitations in the perturbative approach, which are reasonably large. This shows that the triple excitations are important to improve
accuracies of our calculations. Our results are compared with the experimental values listed in the National Institute of Science and
Technology (NIST) database \cite{NIST}. Differences between our final results from these experimental values are given as $\Delta$ in
percentage in the same table. This shows that our calculations agree with the experimental values at the sub-one percentage level in all
the states. These uncertainties can be reduced further by incorporating full triple excitations in our RCC method. Nonetheless, this analysis
shows that we shall be able to obtain $g_j$ factors with similar accuracies for the considered states as these quantities are sensitive to
accuracies of the wave functions in the far nuclear region as for the case of energies.

\begin{table*}[]
\caption{The $g_j$ factors of the $5-7s~^2S_{1/2}$, $5-6p~^2P_{1/2,3/2}$, $5d~^2D_{3/2,5/2}$, and $4f~^2F_{5/2,7/2}$ states. The Dirac
contributions to the $g_j$ factor denoted by $g_j^D$ from the DHF and RCCSD methods along with corrections due to the perturbatively triple
excitation terms and the Breit and QED electronic interactions are given. Contributions to $\Delta g_j^{Q}$ from the DHF and RCCSD method
are quoted correspondingly. The final values, $g_j^D+\Delta g_j^Q$ by adding corrections to $g_j^D$, are taken from the RCCSD method. The
uncertainties are mentioned in the parentheses by estimating as half of the contributions due to the perturbatively triple excitation terms. \label{Cdgj}}
{\setlength{\tabcolsep}{8pt}
\begin{tabular}{l |cc |ccc |cc |l}\hline\hline \\
State &  \multicolumn{2}{c|}{Contributions to $g_j^D$}  &  \multicolumn{3}{c|}{Corrections to $g_j^D$} & \multicolumn{2}{c|}{Contributions to $\Delta g_j^Q$} & Final \\
                           \cline{2-3}  \cline{4-6} \cline{7-8} \\
                &DHF 	  &RCCSD 	    &Triples     & Breit    & QED       & DHF	     & RCCSD     &  \\ \hline
$5s~^2S_{1/2}$	&1.999876 &	2.001624 	&-0.001064	 &-0.000020 &-0.0000050 &0.002320 	 &0.002324 	 &2.00286(53)	\\
$5p~^2P_{1/2}$	&0.666604 &	0.666568 	&0.001662	 &0.000013 	&0.0000007 	&-0.000773 	 &-0.000773  &0.66747(83)	\\
$5p~^2P_{3/2}$	&1.333283 &	1.333522 	&0.000860	 &-0.000003 &-0.0000006 &0.000773 	 &0.000773 	 &1.33515(43)	\\
$6s~^2S_{1/2}$	&1.999967 &	2.000202 	&-0.000253	 &-0.000003 &-0.0000007 &0.002320 	 &0.002321 	 &2.00227(13)	\\
$6p~^2P_{1/2}$	&0.666647 &	0.666657 	&0.000407	 &0.000004 	&0.0000003 	&-0.000773 	 &-0.000773  &0.66630(20)	\\
$6p~^2P_{3/2}$	&1.333315 &	1.333396 	&0.000189	 &-0.000001 &-0.0000003 &0.000773 	 &0.000773 	 &1.33436(1)	\\
$7s~^2S_{1/2}$	&1.999984 &	2.000070 	&-0.000099	 &-0.000001 &-0.0000003 &0.002320 	 &0.002320 	 &2.00229(5)	\\
$5d~^2D_{3/2}$	&0.799982 &	0.800013 	&0.000645	 &0.000002 	&-0.0000001 &-0.000464 	 &-0.000464  &0.80020(32)	\\
$5d~^2D_{5/2}$	&1.199982 &	1.200094 	&0.000375	 &-0.000002 &-0.0000014 &0.000464 	 &0.000464 	 &1.20093(19)	\\
$4f~^2F_{5/2}$	&0.857136 &	0.857140 	&-0.000027	 &0.000000 	&0.0000001 	&-0.000331 	 &-0.000331  &0.85678(1)	\\
$4f~^2F_{7/2}$	&1.142850 &	1.142855 	&-0.0000427	 &0.000000 	&-0.0000001 &0.000331 	 &0.000331 	 &1.14314(2)	\\ \hline\hline
\end{tabular}}
\end{table*}

\begin{table}[t]
\caption{The contributions to $g_j^D$ obtained in the perturbative triples using the RCCSDpT method.} \label{CdgjpT}
{\setlength{\tabcolsep}{2pt}
\begin{tabular}{lrrrr}\hline\hline
States &$T_2^{\dagger}OS_{3v}^{pert}$ &$S_{1v}^{\dagger}T_2^{\dagger}OS_{3v}^{pert}$ &$T_{3}^{pert \dagger}OT_{3}^{pert}$ & $S_{3v}^{pert \dagger}OS_{3v}^{pert}$ \\\hline	
$5s~^2S_{1/2}$ &-0.001070  &-0.000037 &0.000154 &-0.000111 \\
$5p~^2P_{1/2}$ &0.000830   &0.000050  &0.000033 &0.000749  \\
$5p~^2P_{3/2}$ &0.000137   &0.000011  &0.000073 &0.000639  \\
$6s~^2S_{1/2}$ &-0.000286  &-0.000030 &0.000026 &0.000037  \\
$6p~^2P_{1/2}$ &0.000173   &-0.000018 &0.000008 &0.000244  \\
$6p~^2P_{3/2}$ &-0.000001  &-0.000014 &0.000018 &0.000186  \\
$7s~^2S_{1/2}$ &-0.000107  &-0.000011 &0.000010 &0.000009  \\
$5d~^2D_{3/2}$ &0.000025   & $\sim 0.0$  &0.000027 &0.000586\\
$5d~^2D_{3/2}$ &-0.000056  & $\sim 0.0$  &0.000041 &0.000389  \\
$4f~^2F_{5/2}$ &-0.000027  & $\sim0.0 $ & $\sim0.0$ & $\sim 0.0$ \\
$4f~^2F_{7/2}$ &-0.000043  & $\sim 0.0$ & $\sim 0.0$  & $\sim 0.0$ \\\hline\hline
\end{tabular}}
\end{table}

After the investigation of roles of the electron correlation effects in the EAs, we present the calculations of $g_j$ factors of the above
aforementioned states of $^{113}$Cd$^+$. We give the $g_j^D$ values in Table \ref{Cdgj} from the DHF and RCCSD methods along with the corrections
from the terms involving perturbative triple operators of the RCCSDpT method. The estimated $\Delta g_j^Q$ corrections to the $g_j^D$ values
are also quoted from the DHF and RCCSD methods in the same table. They seem to be the decisive contributions to obtain the final results.
Correlation contributions to the $g_j^D$ values due to the perturbative triple excitations, and contributions from the Breit and QED
interactions are shown explicitly. The final $g_j$ values, $g_j=g_j^D+\Delta g_j^Q$, of the respective states are obtained by adding all
these corrections. In our calculations we find the perturbatively triple excitation terms have the sizable contributions to the $g_j$
factors. This suggests full account of triple and other higher level excitations would improve the results further. Nonetheless, we
anticipate additional contributions from these higher level excitations will be within the half of the estimated contributions due to the
perturbative triple excitations. On this basis we assign uncertainties as 50\% of the perturbative excitation contributions to the final
values of the $g_j$ factors. There are no experimental values of the $g_j$ factors of the considered states in Cd$^+$ available to
compare with our calculations.

\begin{figure}[t]
\begin{center}
 \includegraphics[width=8cm]{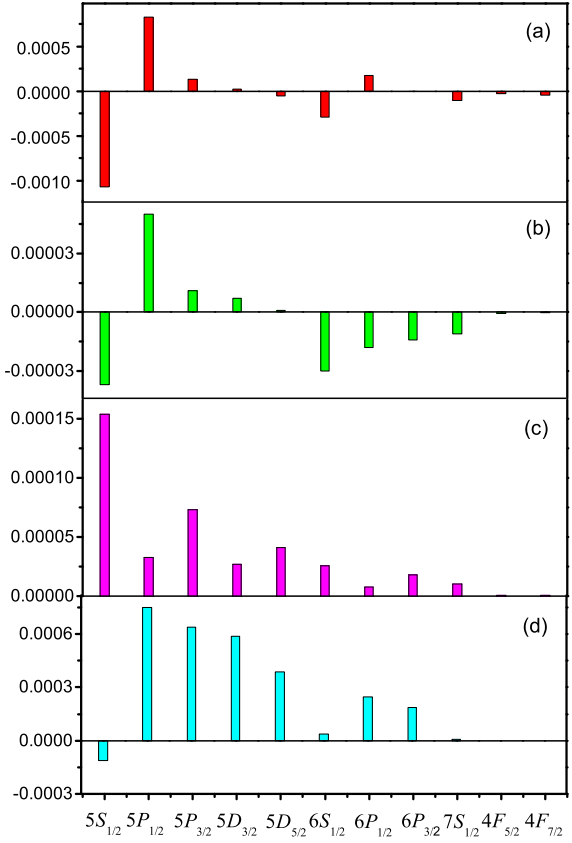}
\end{center}\vspace{-6mm}
\caption{The values of the triple contributions of (a) $T_2^{\dagger}OS_{3v}^{pert}$, (b) $S_{1v}^{\dagger}T_2^{\dagger}OS_{3v}^{pert}$,
(c) $T_{3}^{pert \dagger}OT_{3}^{pert}$, and (d) $S_{3}^{pert \dagger}OS_{3v}^{pert}$ for different valence states.}
\label{figTriple}
\end{figure}

The contributions to $g_j^D$ of the perturbatively triple excitations terms originate mainly from the $T_2^{\dagger}OS_{3v}^{pert}$, $S_{1v}^{\dagger}T_2^{\dagger}OS_{3v}^{pert}$, $T_{3}^{pert
\dagger}OT_{3}^{pert}$, and $S_{3}^{pert \dagger}OS_{3v}^{pert}$ terms , which is given explicitly in Table \ref{CdgjpT}. Computing these terms is very time
consuming. As can be seen that these triple contributions are as large as compared to the correlation contributions due to the RCCSD method.
Contributions from $T_2^{\dagger}OS_{3v}^{pert}$ are found to be relatively larger compared to the other terms followed by the
$T_{3}^{pert \dagger}OT_{3}^{pert}$, $S_{1v}^{\dagger}T_2^{\dagger}OS_{3v}^{pert}$, and $S_{3}^{pert \dagger}OS_{3v}^{pert}$ and there are
strong cancelations among the contributions from the above terms. The computation of $S_{3}^{pert \dagger}OT_{3}^{pert}$ is extremely
costly, and hence it is neglected in the $g_j^D$ calculation of $4f$ states. As shown in \ref{CdgjpT} the contributions of the triple
excitation terms for the $4f$ states are far less than the other states. In Fig. \ref{figTriple}, we plot magnitudes of these contributions
from the individual terms explicitly in order to highlight their roles. It can be seen from these figures that the triples excitations have substantially big contributions for the ground state and $n=5$ excited states and smaller contributions for higher-lying excited states.

Another motivation of the present work is to estimate the typical order of systematic effect due to the applied magnetic field to the
microwave clock frequency of the ground state hyperfine structure splitting in $^{113}$Cd$^+$. The present level of accuracy of this clock
is $\sim 10^{-14}$ \cite{Miaok-OL-2015} and our desired precision level is $10^{-16}$. In $^{113}$Cd$^+$, the $^2S_{1/2}~(F=0, m_F=0)\to (F=1, m_F=0)$
transition is considered for making clock, where $F=I\pm J$ for the nuclear spin $I$ and electron angular momentum $J$ and with its
projection $m_F$. Under the low magnetic field, the energies $W(F,m_F,B)$ of different hyperfine-Zeeman sublevels $|(J,m_J)F,m_F\rangle$
for a given magnetic field strength $B$ can be approximately estimated as \cite{Breit1931,Vanier1989,Itano2001}
\begin{eqnarray}
W(0,0,B)&\simeq& W(0,0,0)- \frac{3hA_{hf}}{4}
          - \frac{[g_j-g_I]^2\mu_B^2B^2}{4hA_{hf}} \label{FMF00} \ \ \ \ \ \ \ \\
W(1,0,B)&\simeq& W(1,0,0)+ \frac{h A_{hf}}{4}
          +\frac{[g_j-g_I]^2\mu_B^2B^2}{4hA_{hf}} , \label{FMF10}\ \ \ \
\end{eqnarray}
and
\begin{eqnarray}
W(1,\pm1,B)&\simeq& W(1,\pm 1,0)+ \frac{h A_{hf}}{4}
          \pm \frac{[g_j+g_I]\mu_B B}{2} \ \ \ \ \ \ \label{FMF11}
\end{eqnarray}
where $g_j$ and $g_I$ are the electronic and nuclear $g$ factors, $\mu_B$ is the Bohr magneton, and $A_{hf}$ is the magnetic dipole hyperfine
structure constant. Using Eqs. (\ref{FMF00}) and (\ref{FMF10}), the energy difference between the clock states of the ground state
in $^{113}$Cd$^+$ is given by
\begin{eqnarray}
W(0,0,B)-W(1,0,B) \simeq -hA_{hf} - \frac{[g_j-g_I]^2\mu_B^2B^2}{2hA_{hf}} . \ \ \ \label{clockEner}
\end{eqnarray}
As can be seen, the energy shift depends on the stray magnetic field, where the second term in Eq. (\ref{clockEner}) is referred to as the
second-order Zeeman shift. Thus, the change in frequency due to the second-order Zeeman shift is expressed as
\begin{equation}
\Delta \nu^{(2)}_{Zeem}(B)=-\frac{[g_j-g_I]^2\mu_B^2B^2}{2h^2A_{hf}}. \label{2ndZeem}
\end{equation}
As can be seen, uncertainty to this quantity depends on the uncertainties in both the $g_j$ factor and the applied magnetic field $B$. In our
experiment, the magnetic field is generated by a pair of Helmholtz coils and the current controls the strength of the magnetic field. Thus,
it is not straightforward for us to measure the applied magnetic field strength very accurately. Under such circumstances, the value of $B$
needs to be calibrated.

Assuming the $B$ value is known precisely, the fractional frequency uncertainty in $\Delta \nu^{(2)}_{Zeem}(B)$ can be obtained by
\begin{eqnarray}
\delta [\Delta \nu^{(2)}_{Zeem}(B)] &=&\frac{\partial (\Delta \nu^{(2)}_{Zeem}(B))}{\partial g_j A_{hf}}\delta g_j \nonumber \\
                                     &=&(g_j-g_I)\frac{\mu_B^2 B^2}{h^2A_{hf}^2}\delta g_j \nonumber \\
                                     &\approx&1.70\times 10^{-14}\delta g_j, \label{Delta Z}
\end{eqnarray}
where $\delta g_j$ is the estimated uncertainty of $g_j$. The above value is estimated by considering $B=10^{-7}$ Tesla for our typical
condition of the experiment, $A_{hf} \simeq-15.2$ GHz \cite{Miaok-OL-2015}, and $g_I$=$0.6223009(9) \times 10^{-3}$ \cite{Spence1972}. By
substituting our determined values $g_j=2.00286$ and $\delta g_j= \pm 0.00053$, we find that uncertainty in the second-order Zeeman shift
will not affect to the clock frequency at the $10^{-16}$ precision level if the applied magnetic field is lower than the above assumed
value. It is also imperative to estimate the maximum level of fluctuation in the applied in order to sustain the aforementioned precision level
of the clock frequency. For this purpose, we use two magnetic-field sensitive transitions, i.e., $\nu_+$ and $\nu_{-}$ correspond to the transitions
from $(F,m_F)=(0,0)$ to $(F,m_{F})=(1,1)$ and from $(F,m_F)=(0,0)$ to $(F,m_{F})=(1,-1)$, respectively. Using
Eqs.(\ref{FMF00}) and (\ref{FMF11}), their energy differences at the first-order level of $B$ can be given by
\begin{eqnarray}
W(0,0,B)-W(1,1,B)&\simeq&-h A_{hf} -\frac{[g_j+g_I]\mu_B B}{2}  \label{v+} \ \ \ \ \
\end{eqnarray}
and
\begin{eqnarray}
W(0,0,B)-W(1,-1,B)\simeq-h A_{hf} +\frac{[g_j+g_I]\mu_B B}{2}.   \label{v-} \ \ \
\end{eqnarray}
This yields frequency difference as
\begin{equation}
\nu_{\mp}\equiv\nu_{-}-\nu_{+} \simeq \frac{[g_j+g_I]\mu_B B}{h} .  \label{vmp}
\end{equation}
Using this expression, the magnetic field strength $B$ can be determined as
\begin{equation}
B\simeq\frac{h\nu_{\mp}}{[g_j+g_I]\mu_B}. \label{B}
\end{equation}
This leads to uncertainty in the calibration of magnetic field with respect to the $g_j$ factor as
\begin{eqnarray}
\delta B&=&\frac{\partial B}{\partial g_j}\delta g_j \simeq -\frac{h\delta \nu_{\mp}}{(g_j+g_I)^2\mu_B}\delta g_j  \nonumber \\
        & \approx & 5.0\times 10^{-8}\delta g_j \ \rm{Tesla}. \label{UB}
\end{eqnarray}
Here, we have used as $\delta \nu_{\mp} \simeq 2.8$ kHz for $B=10^{-7}$ Tesla \cite{Miaok-OL-2015} along with values for other variables as
defined above. According to this, the calibration of the $B$ value will be affected by the uncertainty in the value of $g_j$. Using our
estimated $g_j$ value, we anticipate the uncertainty in $B$ would be less than $10^{-10}$ Tesla. This is sufficiently low to maintain
the uncertainty in the fractional second-order Zeeman shift with respect to the clock frequency lower than  $10^{-17}$ for the assumed
magnetic field. This is good enough for ensuring the $10^{-16}$ precision level measurement of clock frequency.

\section{Conclusion}

The $g_j$ factors of the ground $5s~^2S_{1/2}$ state, and the excited $6-7s~^2S_{1/2}$, $5-6p~^2P_{1/2,3/2}$, $5d~^2D_{3/2,5/2}$ and
$4f~^2F_{5/2,7/2}$ states of $^{113}$Cd$^+$ are calculated by using the relativistic coupled-cluster theory. We have also evaluated
energies of these states and compared them with their experimental results to verify reliability of our calculations. We observed triples
effects are significant for high precision calculation of the $g_j$ factors of the aforementioned states, especially for the low-lying
states, in Cd$^+$. Using the precisely determined $g_j$ factor of the ground state, we have analyzed typical order of systematic shift
due to the second-order Zeeman effect in the clock frequency of the $|5s~^2S_{1/2}, F=0,m_F=0 \rangle \leftrightarrow |5s~^2S_{1/2},
F=1,m_F=0 \rangle$ transition in $^{113}$Cd$^+$ ion and found it will be within the desired fractional uncertainty $10^{-16}$ level.

\section*{Acknowledgement}
We would like to thank V. M. Shabaev and V. A. Dzuba for their helpful suggestions. This work is supported by the National Key Research and Development Program (NKRDP) of China, Grant No. 2016YFA0302100, the National Natural Science Foundation of China under Grant No. 11874064, and the Strategic Priority, and Research Program of the Chinese Academy of Sciences (CAS), Grant No. XDB21030300, and B.K.S. would like to acknowledge use of Vikram-100 HPC of Physical Research Laboratory (PRL), Ahmedabad, and J.Z.H., J.W.Z. and L.J.W. acknowledge the Initiative Program of State Key Laboratory of Precision Measurement Technology and Instruments.

\end{document}